\documentstyle[12pt]{article}
\newcommand{\bm}[1]{\mbox{\boldmath $#1$}}
\newcommand{\rd}{{\rm d}}
\newcommand{\be}{\begin{equation}}
\newcommand{\ee}{\end{equation}}
\newcommand{\ba}{\begin{eqnarray}}
\newcommand{\ea}{\end{eqnarray}}
\newcommand{\bb}[1]{\bibitem{#1}}
\begin{document}
\setcounter{page}{1}
\title{Classical solutions of gravitating Chern-Simons
electrodynamics\thanks{invited talk at the workshop on {\em
Geometry of Constrained Dynamical Systems}, Cambridge 15-18
June 1994}}
\author{G\'erard Cl\'ement
\\
\small Laboratoire de Gravitation et Cosmologie Relativistes\\
\small Universit\'e Pierre et Marie Curie, CNRS/URA769 \\
\small Tour 22-12, Bo\^{\i}te 142 --
 4, place Jussieu, 75252 Paris cedex 05, France}
\date{\small June 27, 1994}
\maketitle
\begin{abstract}
We discuss the reduction of gravitating Chern-Simons
electrodynamics with two commuting Killing vectors to a
dynamical problem with four degrees of freedom, and present
regular particle-like solutions.
\end{abstract}
In three space-dimensions, the abelian Higgs model coupled
to Einstein gravity is known to admit static multi-vortex
solutions \cite{1}. Stationary vortex solutions arise
\cite{2} when a Chern-Simons term for the gauge field
\cite{Sch,DJT} is added. We shall show here that, in this
last case, the Higgs field is not really necessary, as
sourceless gravitating Chern-Simons electrodynamics admits
regular particle-like solutions.

In the present talk, we first discuss the reduction of
gravitating Chern-Simons electrodynamics with two commuting
Killing vectors to a dynamical problem. We then present
exact particle-like solutions to this problem in the case
where the cosmological constant is negative. Finally, we
briefly discuss the generalization to the case where
Einstein gravity is replaced by topologically massive
gravity \cite{DJT}.

The action for gravitating Chern-Simons electrodynamics is
the sum
\be
I = I_G + I_E \, ,
\ee
of the action for Einstein gravity
\be
I_G = -m \, \int \rd ^3 x \, \sqrt{|g|} \, (g^{\mu \nu} \,
R_{\mu \nu} + 2 \, \Lambda)
\ee
($m$ is related to the Einstein gravitational constant
$\kappa$ by $m=1/2\kappa$, while $\Lambda$ is the
cosmological constant), and of the action for Chern-Simons
electrodynamics
\be
I_E = -\frac{1}{4} \, \int \rd ^3 x \, (\sqrt{|g|} \, g^{\mu
\nu} \, g^{\rho \sigma} \, F_{\mu \rho} \, F_{\nu \sigma} -
\mu \, \varepsilon^{\mu \nu \rho} \, F_{\mu \nu} \, A_\rho)
\, ,
\ee
where $F_{\mu \nu}\equiv A_{\nu ; \mu}-A_{\mu ; \nu}$, and
$\mu$ is the topological mass constant.

Let us assume that our three-dimensional space-time has two
commuting Killing vectors $K_1$, $K_2$. The metric $\rd s^2$
is then invariant under the SL(2,R) group of transformations
in the plane $(K_1,K_2)$, which is locally isomorphic to
SO(2,1). A parametrization of the metric making this
invariance manifest is
\be
\rd s^2 = \lambda_{ab}(\rho) \, \rd x^ a \, \rd x^b +
\zeta^{-2}(\rho) \, R^{-2}(\rho) \, \rd \rho^2 \, .
\ee
In (4), $\lambda$ is the $2 \times 2$ matrix
\be
\lambda \equiv \left(
\begin{array}{cc}
T+X & Y \\
Y & T-X
\end{array}
\right)
\ee
of determinant $R^2 \equiv \bm{X}^2$, where
\be
\bm{X}^2 \equiv T^2 - X^2 - Y^2
\ee
is the Minkowski pseudo-norm in the minisuperspace spanned
by the vector $\bm{X}$. The light-cone $R^2=0$ divides this
minisuperspace in three regions. The signature of the metric
(4) is Lorentzian with $\rho$ space-like if $\bm{X}$ is
space-like, Lorentzian with $\rho$ time-like (cosmology) if
$\bm{X}$ is past time-like, and Riemannian if $\bm{X}$ is
future time-like. The function $\zeta (\rho)$ in (4) plays
the role of the usual lapse function by allowing for
arbitrary reparametrizations of the coordinate $\rho$. We
complete this parametrization by assuming the ansatz for the
gauge potentials:
\be
A_\mu \, \rd x^\mu = \psi_a (\rho) \, \rd x^a \, .
\ee

Consider first the case of gravity alone \cite{Cosm}. The
action $I_G$ reduces with the parametrization (4) to
\be
I_G = - \int \rd ^2 x \int \rd \rho \, [\zeta \, \frac{m}{2}
\, \bm{X}^2 + 2 \, \zeta ^{-1} \, m \, \Lambda]
\ee
(with $ \dot{} = \rd / \rd \rho$), where $\zeta$ acts as
Lagrange multiplier. Varying the action (8) with respect to
$\bm{X}$ and fixing $\zeta=1$, we find that the metric
follows a geodesic
\be
\bm{X} = \bm{\alpha} \, \rho + \bm{\beta}
\ee
of minisuperspace. The variation of $I_G$ with respect to
$\zeta$ leads to the Hamiltonian constraint (written for
$\zeta=1$)
\be
H_0 \equiv - \frac{m}{2} \, \dot{\bm{X}}^2 + 2 \, m \,
\Lambda = 0 \, ,
\ee
which fixes the squared slope $\bm{\alpha}^2$. A
particularly interesting class of solutions are the BTZ
black-hole solutions for a negative cosmological constant
$\Lambda=-l^{-2}$ \cite{BTZ},
\ba
\rd s^2 & = & (2 \, l^{-2} \, \rho - \frac{M}{2}) \, \rd t^2
+ J \, \rd t \, \rd \theta - (2 \, \rho + \frac{M \,
l^2}{2}) \, \rd \theta^2 \nonumber \\
& & \mbox{} - [4 \, l^{-2} \, \rho^2 + \frac{J^2 - M^2 \,
l^2}{4}]^{-1} \, \rd \rho^2 \, ,
\ea
where $\theta$ is an angle. These are regular, with two
horizons ($\bm{X}$ past light-like) if $|J| \leq Ml$. The
parameters $M$ and $J$ may be interpreted \cite{BTZ} as the
mass and spin of these `particle-like' metrics.

We now consider the full action (1), which reduces with the
parametrization (4), (7) to $I=\int \rd ^2 x \int \rd \rho
L$, with
\be
L = \frac{1}{2} \, [\zeta \, (-m \, \dot{\bm{X}}^2 +
\dot{\overline{\psi}} \, \bm{\Sigma} \cdot \bm{X} \,
\dot{\psi}) - \mu \, \overline{\psi} \, \dot{\psi} - 4 \,
\zeta^{-1} \, m \, \Lambda] \, .
\ee
The real Dirac-like matrices in (12) are defined by
\be
\Sigma ^0 = \left(
\begin{array}{cc}
0 & 1 \\
-1 & 0
\end{array}
\right) \, , \,\,\,
\Sigma ^1 = \left(
\begin{array}{cc}
0 & -1 \\
-1 & 0
\end{array}
\right) \, , \,\,\,
\Sigma ^2 = \left(
\begin{array}{cc}
1 & 0 \\
0 & -1
\end{array}
\right) \, ,
\ee
and
\be
\overline{\psi} \equiv \psi^T \, \Sigma^0
\ee
is the real adjoint of the `spinor' $\psi$. Our reduced
dynamical system (12) has five degrees of freedom,
parametrized by the coordinates $\bm{X}$, $\psi$ and the
conjugate momenta $\bm{P} \equiv \partial L/ \partial
\dot{\bm{X}}$, $\Pi^T \equiv \partial L/ \partial
\dot{\psi}$. The invariance of the original action under
diffeomorphisms implies the invariance of (12) under SO(2,1)
transformations, which leads to the conservation of the
angular momentum vector
\be
\bm{J} = \bm{L} + \bm{S} \, ,
\ee
sum of `orbital' and `spin' contributions given by
\be
\bm{L} \equiv \bm{X} \wedge \bm{P} \, , \,\,\, \bm{S} \equiv
\frac{1}{2} \, \Pi^T \, \bm{\Sigma} \, \psi \, .
\ee
It follows that, as implied by the notation, the components
$g_{ab}$ of the metric tensor transform vectorially under
the action of SO(2,1), while the gauge potentials transform
spinorially.

In the case $\mu = 0$, the coordinates $\psi_a$ are cyclic,
so that the two corresponding degrees of freedom may be
eliminated altogether \cite{EM}. If $\mu \neq 0$, variation
of the Lagrangian (12) with respect to $\psi$ leads to the
first integrals
\be
\Pi^T - \frac{\mu}{2} \, \overline{\psi} = \varpi^T \, .
\ee
Treating (17) as a pair of second-class constraints, we
eliminate the $\Pi^a$ in terms of the $\psi_a$. Actually it
is more convenient to use, instead of $\psi$, the translated
spinor $\hat{\psi} \equiv \psi - \mu^{-1}
\overline{\varpi}^T$, in terms of which the spin vector
$\bm{S}$ takes the form
\be
\bm{S} = \hat{\bm{S}} + \frac{1}{4 \, \mu} \, \varpi^T \,
\bm{\Sigma} \, \overline{\varpi}^T \, ,
\ee
with
\be
\hat{\bm{S}} = \frac{\mu}{4} \, \hat{\overline{\psi}} \,
\bm{\Sigma} \, \hat{\psi} \, .
\ee
It then follows from the Dirac brackets of the $\hat{\psi}$
(written in matrix form)
\be
[\hat{\psi},\hat{\psi}^T]_D = - \mu^{-1} \, \Sigma^0 \, ,
\ee
that the Dirac algebra of the $\hat{S}^i$ is the
angular-momentum algebra
\be
[\hat{S}^i,\hat{S}^j]_D = \varepsilon^{ijk} \, \hat{S}_k \,
{}.
\ee
Note however that $\hat{\bm{S}}$ is a very special spin
vector, as it is null,
\be
\hat{\bm{S}}^2 = 0 \, ,
\ee
owing to the reality of the $\psi_a$. From now on, we shall
choose the gauge $\varpi^T=0$, which amounts to dropping the
hats in eqs.\  (19)-(22).

The canonical Hamiltonian derived from (12) may be expanded
in terms of the remaining variables $\bm{X}$, $\bm{P}$ and
$\psi$ as
\be
H \equiv - \frac{\bm{P}^2}{2 \, m} - 2 \, \mu \, \bm{S}
\cdot \frac{\bm{X}}{R^2} + 2 \, m \, \Lambda \, ,
\ee
corresponding to a dynamical system with four degrees of
freedom (point particle coupled to a null spin vector). The
energy of this system is fixed by the Hamiltonian constraint
\be
H=0 \, .
\ee

The Hamiltonian (23) generates the equations of motion
\ba
& & \dot{\bm{X}} = - \frac{\bm{P}}{m} \, , \nonumber \\
& & \dot{\bm{P}} = \frac{2 \, \mu}{R^2} \, \bm{S} - \frac{4
\, \mu}{R^4} \, \bm{X} \, (\bm{S} \cdot \bm{X}) \, , \nonumber \\
& & \dot{\bm{S}} = \frac{2 \, \mu}{R^2} \, \bm{X} \wedge
\bm{S} \,\,\,\,\,\, \Longleftrightarrow \,\,\,\,\,\,
\dot{\psi} = \frac{\mu}{R^2} \, \bm{\Sigma} \cdot \bm{X} \,
\psi \, .
\ea
Using these equations, we can easily prove the following
\newline
{\em Theorem}: The only static solution ($Y=0$) to
gravitating Chern-Simons electrodynamics is empty space
($\bm{S}=0$, implying $\psi =0$).

It does not seem possible to obtain the general solution of
eqs.\  (24), (25) in closed form. However we can construct a
class of planar stationary rotationally symmetric solutions
such that $\bm{J} \cdot \bm{X} =0$ for $\Lambda$ negative or
zero (we can show that the constant vector $\bm{J}$ must be
null in this case). These solutions are, for $\Lambda=-l^{-
2}$,
\ba
& & \rd s^2 = (2l^{-2}\rho - \frac{M(\rho)}{2})\rd t^2 -
lM(\rho)\rd t \rd \theta - (2\rho+l^2 \frac{M(\rho)}{2})\rd
\theta^2 - \frac{l^2}{4} \frac{\rd \rho^2}{\rho^2} \, ,
\nonumber \\
& & A_\mu \, \rd x^\mu = c \, \rho^{-\mu l/2} \, (\rd t + l
\, \rd \theta) \, ,
\ea
with
\be
M(\rho) = M - \frac{c^2}{4 \, m} \, \frac{\mu \, l}{\mu \, l
+ 1} \, \rho^{-\mu l} \, .
\ee
For $\mu >0$, $m>0$, $M>0$ the metric (26) is regular ($\rho
=0$ is at infinite geodesic distance), horizonless, and
asymptotic to the BTZ extreme black-hole metric (eq.\  (11)
with $|J|=Ml$). Because of these properties, we suggest that
solution (26) qualifies as the particle-like solution of
gravitating Chern-Simons electrodynamics. In the limit $\mu
\rightarrow 0$, this solution reduces to a dyon extreme
black-hole solution of Einstein-Maxwell gravity \cite{EM}.
The planar `electrostatic' solutions for $\Lambda=0$,
\ba
& & \rd s^2 = (a + b \, r + \frac{c^2}{4 \, m} \, {\rm e}^{-
2\mu r}) \, \rd t^2 - 2 \, \sigma_0 \, \rd t \, \rd \theta -
\rd r^2 \, , \nonumber \\
& & A_\mu \, \rd x^\mu = c \, {\rm e}^{-\mu r} \, \rd t \, ,
\ea
are again regular for $\mu>0$, $m>0$.

We briefly discuss the generalization to the case where
Einstein gravity is replaced by topologically massive
gravity (TMG). The action (8) must be replaced by the action
for TMG
\ba
I_G & = & -m \, \int \rd ^3 x [\sqrt{|g|} \, (g^{\mu \nu} \,
R_{\mu \nu} + 2 \, \Lambda) \nonumber \\
& & \mbox{} -\frac{1}{2 \, \mu'} \, \varepsilon ^{\lambda
\mu \nu} \, {\rm Tr} \, (\Gamma_\lambda \, \partial_\mu
\Gamma _\nu + \frac{2}{3} \, \Gamma_\lambda \, \Gamma_\mu \,
\Gamma_\nu)] \, ,
\ea
where the $\Gamma_\lambda$ are the connections written in
matrix form, and $\mu'$ is the topological mass constant for
gravity. This gives rise, with the parametrization (4), to
the generalized Lagrangian
\be
L_G = - \frac{m}{2} \, [- \frac{1}{\mu'} \, \zeta^2 \,
(\bm{X} \wedge \dot{\bm{X}}) \cdot \ddot{\bm{X}} + \zeta \,
\dot{\bm{X}}^2 + 4 \, \zeta^{-1} \, \Lambda] \, .
\ee
Solutions to the equations of motion deriving from (30) are
discussed in \cite{Part}. The equations of motion for
Chern-Simons electrodynamics coupled to TMG are currently
under investigation \cite{AM}. We have been able to show
that these equations again admit particle-like solutions of
the form (26), with the `mass function' (27) replaced by
\be
M(\rho) = M - a \, \rho^{(1-\mu'l)/2} - b \, \rho^{-\mu l}
\, .
\ee

To conclude, let us mention that extreme BTZ black holes
(just as extreme Reissner-Nordstr\"{o}m black holes) do not
interact, so that stationary multi-extreme-black-hole
systems (with auxiliary conical singularities) are possible
\cite{BH}. It is very likely that stationary multi-particle
solutions to gravitating Chern-Simons electrodynamics may
likewise be constructed from the particle-like solution
(26).

\end{document}